\documentclass
[prb,aps,amssym,superscriptaddress,footinbib,floatfix,twocolumn,notitlepage]{revtex4-1}%
\usepackage{amsfonts}
\usepackage{amsmath}
\usepackage{amssymb}
\usepackage{color}
\usepackage{graphicx}%
\providecommand{\U}[1]{\protect\rule{.1in}{.1in}}

\renewcommand{\vec}[1]{\mathbf{#1}}
\newcommand{\ve}{\vec e}
\newcommand{\vk}{\vec k}
\newcommand{\vq}{\vec q}
\newcommand{\vp}{\vec p}
\renewcommand{\vr}{\vec r}
\newcommand{\vsigma}{\mbox{\boldmath $\sigma$}}
\newcommand{\ketplus}{|+\rangle}

\newcommand{\vkperp}{\vk_{\perp}^{\vphantom{x}}}
\newcommand{\vkperpp}{\vk_{\perp}'}

\begin{document}
\title{Disorder correction to the minimal conductance of a nodal-point semimetal}
\author{Zheng Shi}
\affiliation{Dahlem Center for Complex Quantum Systems and Physics Department, Freie
Universit\"{a}t Berlin, Arnimallee 14, 14195 Berlin, Germany}
\author{Bj\"{o}rn Sbierski}
\affiliation{Dahlem Center for Complex Quantum Systems and Physics Department, Freie
Universit\"{a}t Berlin, Arnimallee 14, 14195 Berlin, Germany}
\affiliation{Department of Physics, University of California, Berkeley, CA 94720, USA}
\author{Piet W. Brouwer}
\affiliation{Dahlem Center for Complex Quantum Systems and Physics Department, Freie
Universit\"{a}t Berlin, Arnimallee 14, 14195 Berlin, Germany}
\date{\today}

\begin{abstract}
We consider the disorder-induced correction to the minimal conductance of an anisotropic two-dimensional Dirac node or a three-dimensional Weyl node. An analytical expression is derived for the correction $\delta G$ to the conductance of a finite-size sample by an arbitrary potential, without taking the disorder average, in second-order perturbation theory. Considering a generic model of a short-range disorder potential, this result is used to compute the probability distribution $P(\delta G)$, which is compared to the numerically exact distribution obtained using the scattering matrix approach. We show that $P(\delta G)$ is Gaussian when the sample has a large width-to-length ratio, and study how the expectation value, the standard deviation, and the probability of finding $\delta G < 0$ depend on the anisotropy of the dispersion.
\end{abstract}
\maketitle

\section{Introduction\label{sec:intro}}

Dirac materials, such as the two-dimensional graphene\cite{Nature.438.197,*RevModPhys.81.109} and three-dimensional Weyl semimetals,\cite{RevModPhys.90.015001} have been a continued focus of research in contemporary condensed matter physics. In these materials conduction and valence bands touch and disperse linearly at discrete ``nodal'' points in reciprocal space.
In the absence of parallel conduction channels, the conductivity of a Dirac semimetal has a characteristic minimum if the Fermi energy is at the nodal point, which reflects the vanishing density of states at this energy. In the absence of disorder, the minimum conductivity of graphene is theoretically predicted to take the universal value $\sigma_{\rm min} = e^2/\pi h$ per valley and per spin direction\cite{PhysRevLett.96.246802,PhysRevB.92.205408} (the condition of approaching universality is further explained below), whereas $\sigma_{\rm min} = 0$ for a Weyl semimetal at zero temperature. \cite{PhysRevB.89.035410,PhysRevLett.113.026602} The presence of disorder that is sufficiently smooth and does not induce inter-node scattering increases $\sigma_{\rm min}$. \cite{PhysRevLett.99.106801,PhysRevB.76.195445,PhysRevB.79.201404,PhysRevLett.113.026602,PhysRevB.92.115145}

\begin{figure}
\centering
\includegraphics[width=0.3\textwidth]{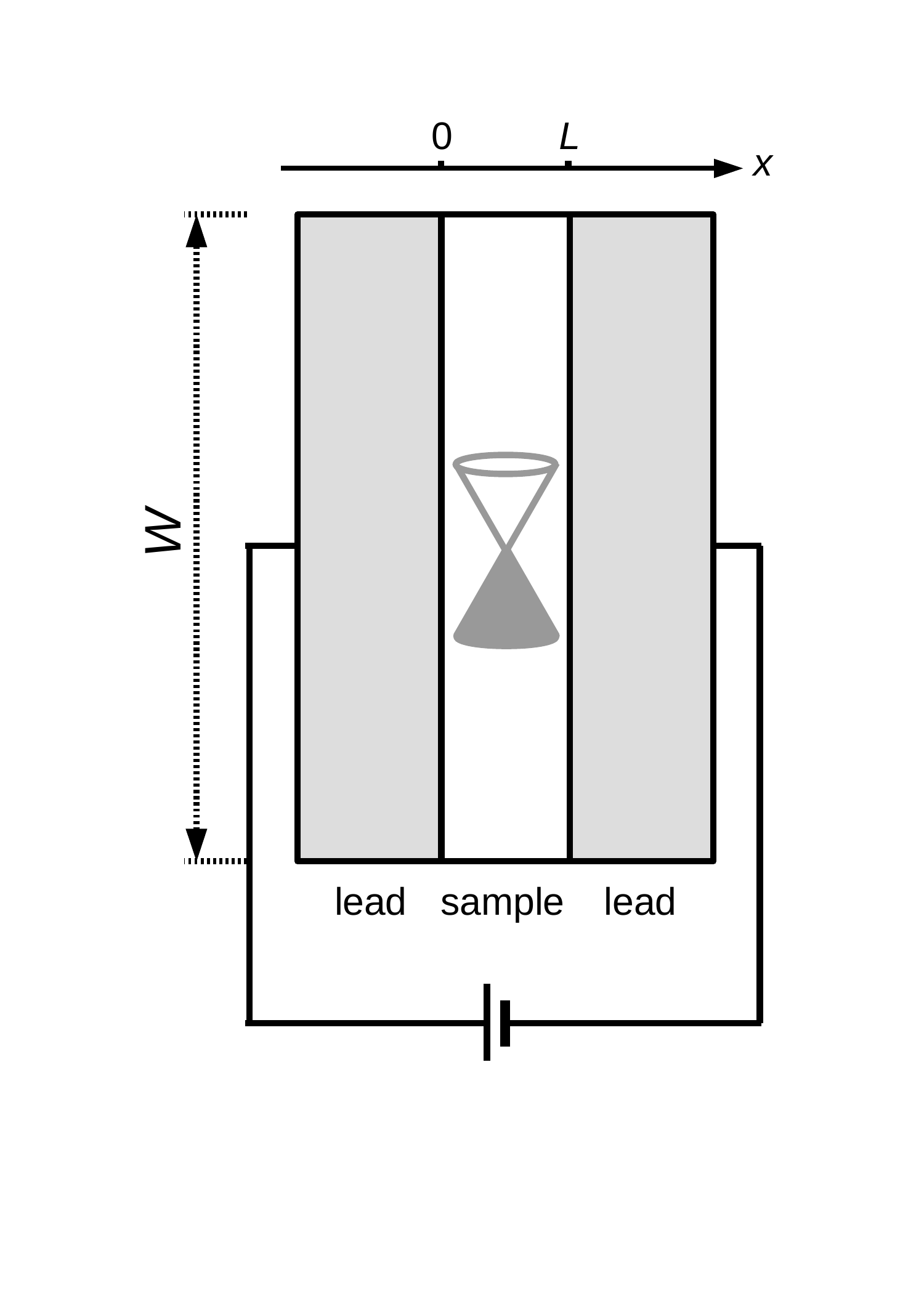}
\caption{Schematic of the geometry considered in this work. The graphene (or Weyl semimetal) sample of length $L$ and width $W$ (or cross section $W^2$) is sandwiched between two highly conductive leads. The Fermi energy is at the nodal point inside the sample.
\label{fig:setup}}
\end{figure}

An experimental conductivity measurement, as well as a theoretical calculation using the Landauer-B\"uttiker approach, involves systems of finite size and addresses the conductance $G$ rather than the conductivity $\sigma$. An idealized geometry consists of a graphene sheet of width $W$ and length $L$ with the Fermi energy near the Dirac point (or a Weyl semimetal of cross section $W^2$ and length $L$), coupled to source and drain reservoirs, see Fig.~\ref{fig:setup} for a schematic picture. The reservoirs consist of a highly doped and hence highly conductive version of the same material. \cite{PhysRevLett.96.246802} Without disorder, neglecting interactions, and with the Fermi energy precisely at the nodal point, for a two-dimensional Dirac semimetal one then has the minimal {\em conductance}\cite{PhysRevLett.96.246802,PhysRevB.92.205408}
\begin{equation}
  G_{{\rm min},\, 2d} = \frac{e^2}{h}\, \frac{W}{\pi L} \label{eq:Gmin2}
\end{equation}
per node, which can be easily translated to a conductivity using the relation $G = \sigma W/L$. In three dimensions one has \cite{PhysRevB.89.035410,PhysRevLett.113.026602}
\begin{equation}
  G_{{\rm min},\, 3d} = \frac{e^2}{h}\, \frac{W^2 \ln 2}{2 \pi L^2}, \label{eq:Gmin3}
\end{equation}
which corresponds to $\sigma_{\rm min} = 0$ in the thermodynamic limit $W$, $L \to \infty$. The derivation of Eqs.~(\ref{eq:Gmin2}) and (\ref{eq:Gmin3}) requires that $W \gg L$ and assumes that the dispersion at the nodal points is isotropic and has no tilt. In the opposite limit $W \lesssim L$ the conductance $G$ depends on the transverse boundary conditions and no universal minimum conductance can be obtained. \cite{PhysRevLett.96.246802} If the Fermi energy is not at the nodal point, $G$ is proportional to $(k_{\rm F} W)^{d-1}$ in the absence of disorder and interactions, with $k_{\rm F} \neq 0$ the Fermi wavevector measured from the nodal point, and $d$ the system dimension, which is a hallmark of ballistic transport with a formally infinite conductivity.


Numerical and analytical calculations for an isotropic nodal point in two or three dimensions show that the {\em average} value of the conductance at $k_{\rm F} = 0$ increases in the presence of disorder, \cite{PhysRevB.79.075405} whereas the conductivity becomes finite for $k_{\rm F} \neq 0$. Here, the disorder is taken to be smooth enough that it scatters carriers within a node only, whereas inter-node scattering remains suppressed. 
In two dimensions, the disorder-induced increase of the conductance at the nodal point leads to an increase of the minimum conductivity, consistent with scaling theory, \cite{PhysRevLett.99.106801,PhysRevB.76.195445,PhysRevB.79.201404} whereas in three dimensions the conductivity at the nodal point is believed to remain zero up to a critical disorder strength. \cite{PhysRevLett.113.026602,PhysRevB.90.241112,PhysRevLett.115.246603,SciRep.6.32446,PhysRevLett.116.066401,PhysRevB.92.115145,PhysRevB.93.075108,PhysRevB.93.155113,PhysRevB.96.064203} (Rare disorder fluctuations may, however, lead to a small residual conductivity even at weak disorder strengths. \cite{PhysRevB.89.245110,PhysRevLett.121.215301})

In this article we address the full probability distribution of the nodal-point conductance in the geometry of Fig.~\ref{fig:setup}, focusing on the regime of weak disorder, in which $G$ is still numerically close to the quasi-ballistic limits Eq.~(\ref{eq:Gmin2}) or (\ref{eq:Gmin3}), without the assumption that the dispersion is isotropic. (We do not consider tilted dispersions, \cite{kobayashi2007,PhysRevB.78.045415,PhysRevB.91.115135,soluyanov2015,PhysRevB.95.045139} however.) The standard diagrammatic theory of conductance fluctuations, which predicts a universal value for the second moment of the distribution depending only on symmetry and dimensions, \cite{PhysRevB.77.193403,PhysRevB.78.033404} is not applicable at the nodal point $k_{\rm F} = 0$, since it requires the limit of $k_{\rm F}$ large compared to the inverse mean-free path. Nevertheless, the same universal value has been reported in numerical studies at the Dirac point in strongly disordered graphene, \cite{EurophysLett.79.57003,PhysRevLett.109.096801} and also near the nodal point in the diffusive phase of the Weyl semimetal, \cite{PhysRevB.96.134201} until impurity scattering between the Weyl nodes causes them to annihilate. In comparison, the quasi-ballistic regime of weak disorder, which we consider here, has received less attention. Analytical and numerical results exist for the average conductance in the isotropic case, \cite{PhysRevB.79.075405} as well as for the disorder-averaged full-counting statistics in two dimensions, \cite{PhysRevB.82.085419} but it is not well understood how the nodal-point conductance fluctuates across different disorder realizations.

\begin{figure}
[h]\centering\includegraphics[width=0.4\textwidth]{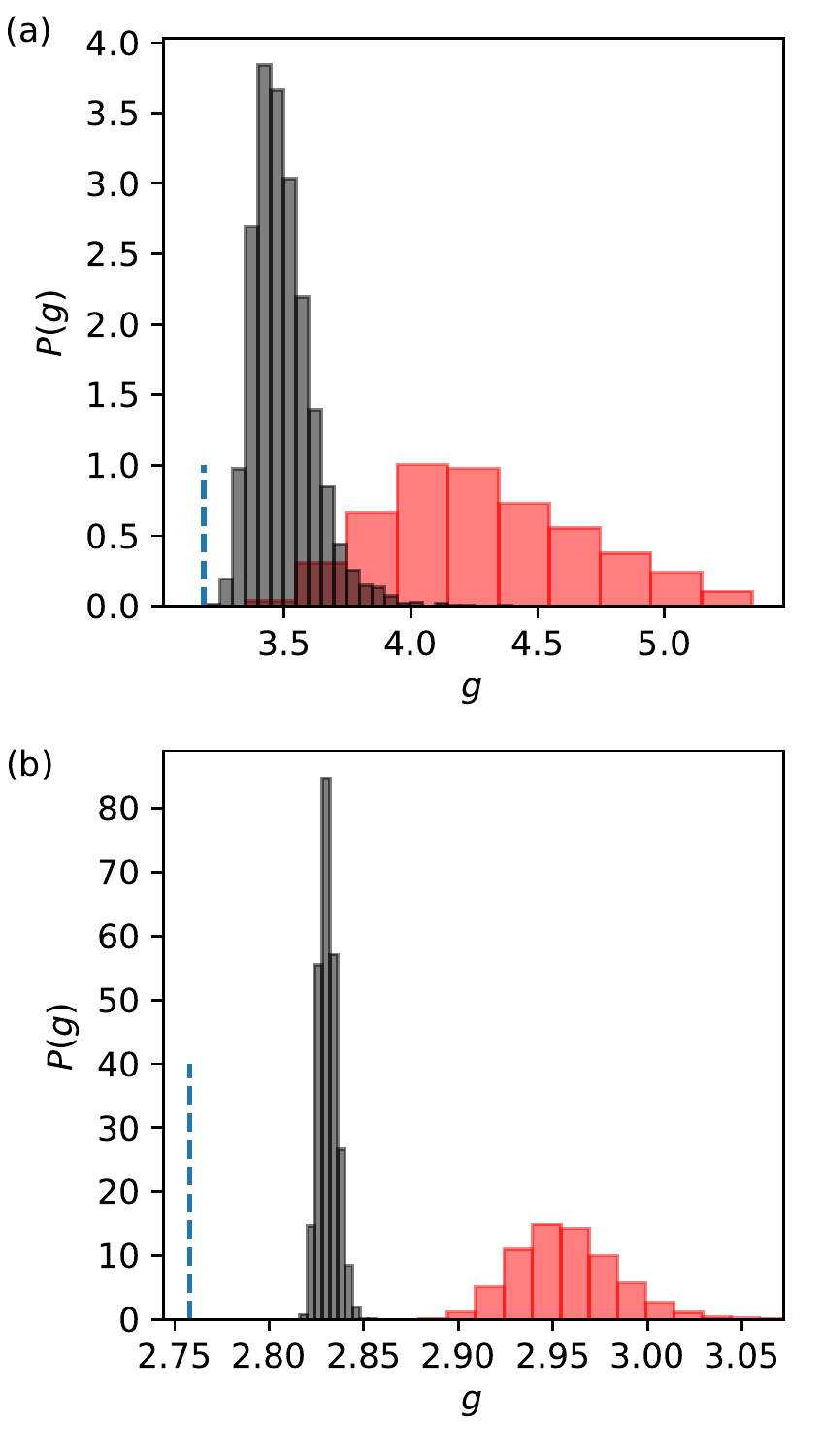}
\caption{Numerical calculation of the distribution of the conductance $G = (e^2/h) g$ of a two-dimensional Dirac node (a) and a three-dimensional Weyl node (b) with an isotropic dispersion and isotropic short-range disorder. The dimensionless disorder strength $K$ and the number $N$ of disorder realizations are $K=1$, $N=9696$ (red) and $K=0.5$, $N=14827$ (gray) in (a) and $K=2$, $N=3683$ (red) and $K=0.5$, $N=3572$ (gray) in (b). The aspect ratio and system size ratio are $W/L = 10$, $L/\xi=20$ in (a) and $W/L = 5$, $L/\xi = 12$ (b), where $\xi$ is the correlation length of the random potential, see Eq.\ (\ref{Urandomsr}) for the definitions of $K$ and $\xi$. The conductances $G_{{\rm min},2d}$ and $G_{{\rm min},3d}$ of quasi-ballistic systems are represented by dashed blue lines. Numerical calculations are performed using the numerical scattering approach of Refs.~\onlinecite{PhysRevLett.99.106801,PhysRevLett.113.026602}. For the data shown here, there is not a single disorder realization with $G < G_{\rm min}$.
\label{fig:finiteK}}
\end{figure}

What motivated this study in particular is the question whether the presence of isotropic short-range disorder may actually lead to a nodal-point conductance that is {\em less} than the quasi-ballistic minimal conductance of Eqs.~(\ref{eq:Gmin2}) and (\ref{eq:Gmin3}). Based on numerical simulations for an isotropic three-dimensional Weyl semimetal, two of us, together with Bergholtz, had conjectured that this does not happen. \cite{PhysRevB.92.115145} (The result of numerical simulations similar to those of Ref.\ \onlinecite{PhysRevB.92.115145} is shown in Fig.\ \ref{fig:finiteK} for a two-dimensional Dirac node and for a three-dimensional Weyl semimetal.) The perturbative analysis we present here show that this conjecture was not correct, although, as we show below, for an isotropic dispersion the probability of such a rare disorder fluctuation is so small that it is not expected to show up in numerical simulations in the parameter regime $W \gtrsim L$ in which the conductance no longer depends on the transverse boundary conditions. When the dispersion itself becomes anisotropic, the probability of the nodal-point conductance being less than the minimal conductance can increase significantly.

The remainder of this article is organized as follows: In Sec.~\ref{sec:model} we briefly review the derivation of Eqs.~(\ref{eq:Gmin2}) and (\ref{eq:Gmin3}) using the Landauer-B\"uttiker approach for the geometry of Fig.~\ref{fig:setup} and their generalization to an anisotropic dispersion. Section~\ref{sec:pert} extends the calculation of the conductance at the nodal point to the second order perturbation theory in a scattering potential. Using a generic model of short-range, isotropic disorder, Sec.~\ref{sec:minimal} then uses the results of Sec.~\ref{sec:pert} to obtain the conductance distribution in perturbation theory, which we compare with the numerically exact results obtained using the scattering matrix approach of Refs.\ \onlinecite{PhysRevLett.99.106801,PhysRevLett.113.026602}. In particular, we calculate the probability that the presence of disorder leads to a decrease of the nodal-point conductance below the quasi-ballistic values of Eqs.~(\ref{eq:Gmin2}) and (\ref{eq:Gmin3}). We conclude in Sec.~\ref{sec:conclusion}.

\section{Conductance in the quasi-ballistic limit\label{sec:model}}

Following Refs.~\onlinecite{PhysRevLett.96.246802,PhysRevLett.113.026602}, we consider a junction consisting of a $d$-dimensional Dirac material, connected to ideal leads. The junction has width $W$ and length $L$, see Fig.~\ref{fig:setup}. We neglect inter-node scattering and allow for an anisotropic dispersion at the nodal point, where we assume that the current flow is along one of the principal axes. With these assumptions, the junction is described at low energies by the Dirac Hamiltonian
\begin{equation}
  H = 
\left\{ \begin{array}{ll}
  v_x p_x \sigma_x + v_y p_y \sigma_y, & d=2, \\
  v_x p_x \sigma_x + v_y p_y \sigma_y + v_z p_z \sigma_z
, & d=3,
  \end{array} \right.
  \label{eq:Hideal}
\end{equation}
for $0 < x < L$, where $v_{x,y,z}$ are the Fermi velocities along the three principal directions, $\vsigma$ is the vector of Pauli matrices, and we take periodic boundary conditions in the transverse directions. To calculate the minimal conductance of a quasi-ballistic junction, the junction is connected to ideal leads for $x < 0$ and $x > L$, which are described by the same Hamiltonian, but with $v_y = v_z = 0$.

Flux-normalized scattering states of the Hamiltonian Eq.~(\ref{eq:Hideal}) that are incident from the left are of the form
\begin{align}
  \psi_{\vkperp}^{\rm in}(\vr) =&\, \frac{e^{i \vkperp \cdot \vr_{\perp}}}{\sqrt{v_x W^{d-1}}} t_{\vkperp}^{(0)} e^{\kappa_{\perp} (x-L)} \ketplus,
  \label{eq:psi_in}
\end{align}
for $0 \le x \le L$, where
\begin{equation}
  t_{\vkperp}^{(0)} = \frac{1}{\cosh \bar k_{\perp} L},\ \
  \bar k_{\perp} = \sqrt{\frac{k_y^2 v_y^2 + k_z^2 v_z^2}{v_x^2}}
  \label{eq:t}
\end{equation}
is the transmission amplitude of the quasi-ballistic junction. Further we abbreviated $\vr_{\perp} = y \ve_y$ for $d=2$ and $\vr_{\perp} = y \ve_y + z \ve_z$ for $d=3$,
\begin{equation}
  \kappa_{\perp} = \left\{ \begin{array}{ll}
   v_y k_y \sigma_z/v_x, & d=2, \\
   (v_y k_y \sigma_z - v_z k_z \sigma_y)/v_x, & d=3,
  \end{array} \right.
  \label{eq:kappa}
\end{equation}
and
\begin{equation}
  |\pm \rangle = \frac{1}{\sqrt{2}} \begin{pmatrix} 1 \\ \pm 1 \end{pmatrix}.
\end{equation}
The transverse wavevector $\vkperp = k_{y} \ve_{y}$ ($\vkperp = k_{y} \ve_{y} + k_z \ve_{z}$) has components $k_{y,z} = 2 \pi m_{y,z}/W$ for $d=2$ ($d=3$), with $m_y$ and $m_z$ integer. 

From the Landauer formula, one then obtains the dimensionless minimal conductance
\begin{equation}
  g_{\rm min} \equiv \frac{h}{e^2}   G_{\rm min} =  \sum_{\vkperp} |t_{\vkperp}|^2
\end{equation}
per node. In general, $g_{\rm min}$ depends on $L$ and $W$ and on the choice of the boundary conditions in the transverse direction. The dependence on the boundary conditions disappears and the dependence on $L$ and $W$ simplifies in the limit $W \gg L$, for which one finds the minimal conductance \cite{PhysRevLett.96.246802,PhysRevB.89.035410,PhysRevB.91.115135} 
\begin{align}
  g_{{\rm min},\, 2d} =&\, \frac{W v_x}{\pi L v_y}, \label{eq:Gmin2anis} \\
  g_{{\rm min},\, 3d} =&\, \frac{W^2 v_x^2 \ln 2}{2 \pi L^2 v_y v_z}, \label{eq:Gmin3anis}
\end{align}
per node, which simplifies to Eqs.~(\ref{eq:Gmin2}) and (\ref{eq:Gmin3}) in the limit of an isotropic dispersion, $v_x = v_y = v_z = v$.

\section{Perturbation theory\label{sec:pert}}

We now consider a potential $U(\vr)$ with support $0 < x < L$, and calculate its effect on the transmission matrix using standard perturbation theory:
\begin{align}
  t_{\vkperp,\vkperpp}
  =&\, t_{\vkperp}^{(0)} \delta_{\vkperp,\vkperpp}
  + \delta t_{\vkperp,\vkperpp}^{(1)} 
  + \delta t_{\vkperp,\vkperpp}^{(2)} + \ldots
\end{align}
where $t_{\vkperp}^{(0)}$ is given by Eq.~(\ref{eq:t}). The leading order correction reads
\begin{equation}
  \delta t_{\vkperp,\vkperpp}^{(1)} = - \frac{i}{\hbar}
  \langle \psi_{\vkperp}^{\rm out}|U|\psi_{\vkperpp}^{\rm in}\rangle,
\end{equation}
where $\psi_{\vkperp}^{\rm in}$ is a flux-normalized scattering state with incoming-wave boundary conditions, see Eq.~(\ref{eq:psi_in}), and $\psi_{\vkperp}^{\rm out}$ is a flux-normalized scattering state with outgoing wave boundary conditions,
\begin{align}
  \psi_{\vkperp}^{\rm out}(\vr) =&\, \frac{e^{i \vkperp \cdot \vr_{\perp}}}{\sqrt{v_x W^{d-1}}} t_{\vkperp}^{(0)} e^{\kappa_{\perp} x} \ketplus
  \label{eq:psi_out}
\end{align}
for $0 \le x \le L$.
The second-order correction is
\begin{equation}
  \delta t^{(2)}_{\vkperp,\vkperpp} =
  - \frac{i}{\hbar} \langle \psi_{\vkperp}^{\rm out}|U G U|\psi_{\vkperpp}^{\rm in}\rangle,
\end{equation}
where $G_{\vkperp}(x,x')$ is the Green function of the quasi-ballistic junction,
\begin{align}
  G_{\vkperp}(x,x') =&\,
  -\frac{it_{\vkperp}^{(0)}}{\hbar v_x} 
  \left\{ \begin{array}{ll}
  e^{\kappa_{\perp} x} |-\rangle \langle - | e^{\kappa_{\perp} (x'-L)}, & \mbox{$x < x'$}, \\
  e^{\kappa_{\perp} (x-L)} |+\rangle \langle + | e^{\kappa_{\perp} x'}, & \mbox{$x > x'$}. \end{array} \right.
\end{align}

Explicitly, we thus find
\begin{widetext}
\begin{align}
  \delta t^{(1)}_{\vkperp,\vkperpp} =&\,
  - \frac{i}{\hbar v_x} t_{\vkperp}^{(0)} t_{\vkperpp}^{(0)} \int_0^L dx
  U_{\vkperpp-\vkperp}(x)
  \langle + | e^{\kappa_{\perp} x} e^{\kappa_{\perp}' (x-L)} | + \rangle, \\
  \delta t^{(2)}_{\vkperp,\vkperp} =&\,
  \frac{1}{\hbar^2 v_x^2} (t_{\vkperp}^{(0)})^2 t_{\vkperpp}^{(0)}
  \int_0^L dx dx'
  U_{\vkperpp-\vkperp}(x) U_{\vkperp-\vkperpp}(x')
  \nonumber \\ &\, \mbox{} \times
  \left[
  \theta(x-x') \langle + | e^{\kappa_{\perp} x} e^{\kappa_{\perp}' (x-L)} | + \rangle \langle + | e^{\kappa_{\perp}'x'} e^{\kappa_{\perp} (x'-L)} | + \rangle
  \right. \nonumber \\ &\, \left. \mbox{} -
 \theta(x'-x) \langle + | e^{\kappa_{\perp} x} e^{\kappa_{\perp}' x} | - \rangle \langle - | e^{\kappa_{\perp}'(x'-L)} e^{\kappa_{\perp} (x'-L)} | + \rangle
  \right], 
\end{align}
\end{widetext}
where 
\begin{equation}
  U_{\vq}(x) = \frac{1}{W^{d-1}}
  \int d\vr_{\perp} e^{i \vq \cdot \vr_{\perp}} U(\vr).
\end{equation}
These equations can then be inserted into the Landauer formula to obtain the conductance. In the following we present the results for $d=2$ and $d=3$ separately.


\subsection{Dirac node}

In the Dirac node case $d=2$, the second-order correction to the conductance $g^{(2)}$ has a particularly simple
form in the large aspect ratio limit $W/L\gg1$:%
\begin{align}
  g^{\left(2\right)} =&\,
  \frac{v_{x}^{2}}{2\hbar^2 v_{y}^{4} L^{4}}
  \int_{0}^{L}dxdx^{\prime}
  \int dydy^{\prime}
  U\left(  \mathbf{r}\right) 
  U\left(\mathbf{r}^{\prime}\right)
  \nonumber \\ &\, \mbox{} \times
  \sum_{\pm}
  \frac{ \pm (  y-y')^{2}}
  {\cosh\frac{\pi v_{x} (y-y')}{v_{y}L} -
     \cos\frac{\pi(x \mp x')}{L}}.
  \label{OUsqcont}%
\end{align}
An alternative expression, which does not require the limit $W/L \gg 1$, can be obtained by expanding the impurity potential $U(\vr)$ as a sine series in the longitudinal ($x$) direction,
\begin{equation}
  U(\vr) = \sum_{\vq} U(\vq)
  e^{-i \vq_{\perp} \cdot \vr_{\perp}} \sin q_{x}x,
  \label{Uxyg}%
\end{equation}
where $q_{x}=\pi m_x /L$, $m_x$ being a positive integer. The Fourier components contribute independently to the conductance at the second order,
\begin{equation}
  g^{\left(2\right)} =
  \frac{1}{4\hbar^{2}v_{x}v_{y}}\frac{W}{L}
  \sum_{\mathbf{q}}\frac{\pi^{2}}{q_{x}^{2}}
  F\left(  \frac{q_{x} L}{\pi},\frac{\pi v_{y}q_{y}}{v_{x}q_{x}}\right)  \left\vert U\left(
\mathbf{q}\right)  \right\vert ^{2}\text{,} \label{G2g}%
\end{equation}
where the dimensionless Fourier coefficients are
\begin{align}
  F
  =&\, \frac{v_{y}q_{x}^{2}L^{2}}{v_x \pi^{2}W}
  \sum_{k_{y}} \sum_{\pm}
  \frac{(\bar k_y + \bar k_y^{\pm})
  \sinh[(\bar k_{y} + \bar k_{y}^{\pm}) L]}
  {(\bar k_y + \bar k_y^{\pm})^2 + q_{x}^{2}}
  \nonumber \\ &\, \mbox{} \times
  \frac{1+\sinh(\bar k_y L)
  \sinh(\bar k_y^{\pm} L)
  \cosh[(\bar k_y - \bar k_y^{\pm})L]}
  {\cosh^{3} (\bar k_y L)
  \cosh^{3} (\bar k_y^{\pm} L)}.
  \label{Fgmq}
\end{align}
Here $\bar k_y$ was defined in Eq.\ (\ref{eq:t}) and we abbreviated
\begin{equation}
  k_y^{\pm} = k_y \pm q_y,\ \
  \bar k_y^{\pm} = \frac{k_y^{\pm} v_y}{v_x}.
\end{equation}
Note that the first argument of the Fourier coefficient $F$ is the integer $m_x$ that labels $q_{x}$. The Fourier component $F(m_x,t)$, with $t = \pi v_y q_y/v_x q_x$, has a well-defined limit when $W/L\rightarrow\infty$ and either $m_x \rightarrow\infty$ or $t \to \infty$,
\begin{equation}
  \left.  F(  m_x,t)  \right\vert _{W/L\to\infty;\, m_x\text{ or } t\to\infty}=
  \frac{4}{\pi}\frac{\pi^{2}-t^{2}}{\left(  \pi^{2}+t^{2}\right)  ^{2}}
  \text{.} \label{sigma2g}%
\end{equation}

\subsection{Weyl node}

In three dimensions there is no closed-form expression for $g^{(2)}$ using the real-space representation of the potential $U$. Instead, expanding the potential $U(\vr)$ as in Eq.\ (\ref{Uxyg}), we find
\begin{align}
  g^{\left(2\right)}=&\, \frac{1}{4\hbar^{2}v_y v_z}\frac{W^{2}}{L^{2}}\sum_{\mathbf{q}%
}\frac{\pi^{2}}{q_{x}^{2}}\left\vert U\left( \mathbf{q}\right)  \right\vert ^{2}  \nonumber\\ &\, \ \ \ \ \mbox{} \times 
  F\left(  \frac{q_{x}L}{\pi},\frac{\pi
v_{y}q_{y}}{v_{x}q_{x}},\frac{\pi v_{z}q_{z}}{v_{x}q_{x}}\right)  \text{,} \label{G2W}%
\end{align}
where the dimensionless Fourier coefficients $F$ read
\begin{widetext}
\begin{align}
  F =&\,
  \frac{q_{x}^{2}L^{2}}{8 \pi^{2}}
  \frac{L v_{y} v_{z}}{W^{2}v_{x}^{2}}
  \sum_{\mathbf{k}_{\perp}} \sum_{\pm}
  \frac{(  \bar{k}_{\perp} \pm \bar{k}_{\perp}^{\pm })
    \sinh[(\bar{k}_{\perp} \pm \bar{k}_{\perp}^{\pm}) L]}
   {[q_{x}^{2} + (\bar{k}_{\perp} \pm \bar{k}_{\perp}^{\pm})^2]
  \cosh^{3}(\bar{k}_{\perp} L) \cosh^{3}(\bar{k}_{\perp}^{\pm}L) }
  \left(  1 \pm \frac{v_{y}^{2} k_{y} k_{y}^{\pm}
  + v_{z}^{2}k_{z} k_{z}^{\pm}}
    {v_{x}^{2}\bar{k}_{\perp}\bar{k}_{\perp}^{\pm}}
  \right)
  \nonumber\\ &\, \ \ \ \ \mbox{}  \times
  \left\{ 3 + \cosh(2 \bar k_{\perp} L) + \cosh(2 \bar k_{\perp}^{\pm} L)
  - \cosh[2(\bar k_{\perp} \mp \bar k_{\perp}^{\pm}) L] \right\}
   .
\label{FWmq1}
\end{align}
\end{widetext}
The rescaled transverse momentum $\bar{k}_{\perp}$ was defined in Eq.\ (\ref{eq:t}). We further abbreviated $k_{y,z}^{\pm} = k_{y,z} \pm q_{y,z}$ and 
\begin{equation}
  \bar k_{\perp}^{\pm} = \frac{1}{v_x}
  \sqrt {v_{y}^{2} (k_{y} \pm q_y)^{2}
  + v_{z}^{2} (k_{z} \pm q_z)^2}.
\end{equation}
In the limit $W/L\gg1$, the dependence on the transverse momentum $\vq_{\perp}$ is through the rescaled magnitude 
\begin{equation}
  \bar q_{\perp} = \sqrt{\frac{v_y^2 q_y^2 + v_z^2 q_z^2}{v_x^2}}
  \label{qbarperp}
\end{equation}
only. Defining $t = \pi \bar q_{\perp}/q_x$, then if one also takes the limit $m_x \to \infty$ or $t\to\infty$ one finds the simple limiting value
\begin{equation}
\left.  F( m_x,t) \right|_{W/L\to\infty;\, m_x\text{ or }t\to\infty}
  =  \frac{2\ln2}{\pi}\frac{2\pi^{2}-t^{2}}
  {\left(  \pi^{2}+t^{2}\right)  ^{2}}.
  \label{sigma2W}%
\end{equation}

It is worth mentioning that Eqs.~(\ref{sigma2g}) and (\ref{sigma2W}) are
consistent with previous results for anisotropic Dirac or Weyl
nodes. \cite{PhysRevB.91.115135} In the case of an isotropic two-dimensional Dirac node 
with $v_x=v_y=v$, the periodic potential $U=U_{0}\sin(q_{x}x) \cos
(q_{y}y)$ couples the states labeled by momentum $\vp = p_x \ve_x + p_y \ve_y$
to those labeled by momenta $\vp + s_x q_x \ve_x + s_y q_y \ve_y$, with $s_{x,y} = \pm 1$.
Using polar coordinates, $\vp = p (\cos \theta \ve_x + \sin \theta \ve_y)$ and 
$\vq = q (\cos \phi \ve_x + \sin \phi \ve_y)$, with $q^2 = q_x^2+q_y^2$, for $p\ll q$ there is an
anisotropic correction $\delta v ( \theta)$ to the Fermi velocity
$v$, defined by the dispersion relation $\varepsilon(\vp) = [v + \delta v(\theta)] p$.
Up to second order in $U$, we find
\begin{align}
  \delta v(  \theta)  =&\,
  -\frac{U_{0}^{2}}{2\hbar^2 v q^{2}}\sin^{2}(\theta-\phi).
\end{align}
Combining this velocity renormalization with the expression (\ref{eq:Gmin2anis}) for the conductivity for an anisotropic Dirac node, \cite{PhysRevB.91.115135} one immediately reproduces Eq.~(\ref{sigma2g}),%
\begin{align}
  g =&\, \frac{W}{L}\frac{1}{\pi}\frac{v+\delta v\left(  0\right)  }{v+\delta
v\left(  \frac{\pi}{2}\right)  }
  \nonumber \\ \approx&\, \frac{W}{L}\frac{1}{\pi}\left(
  1+\frac{q_{x}^{2}-q_{y}^{2}}{q^4}%
  \frac{U_{0}^{2}}{2\hbar^2 v^{2}}\right).
\end{align}
The expression in the case of a three-dimensional Weyl node, Eq.~(\ref{sigma2W}), is similarly reproduced.

The perturbation-theory expressions Eqs.~(\ref{OUsqcont}), (\ref{G2g}) and
(\ref{G2W}) constitute the central results of this work. We emphasize that they are
valid for an arbitrary weak potential $U$ and do not involve a disorder average.

\begin{figure}
[h]\centering\includegraphics[width=0.4\textwidth]{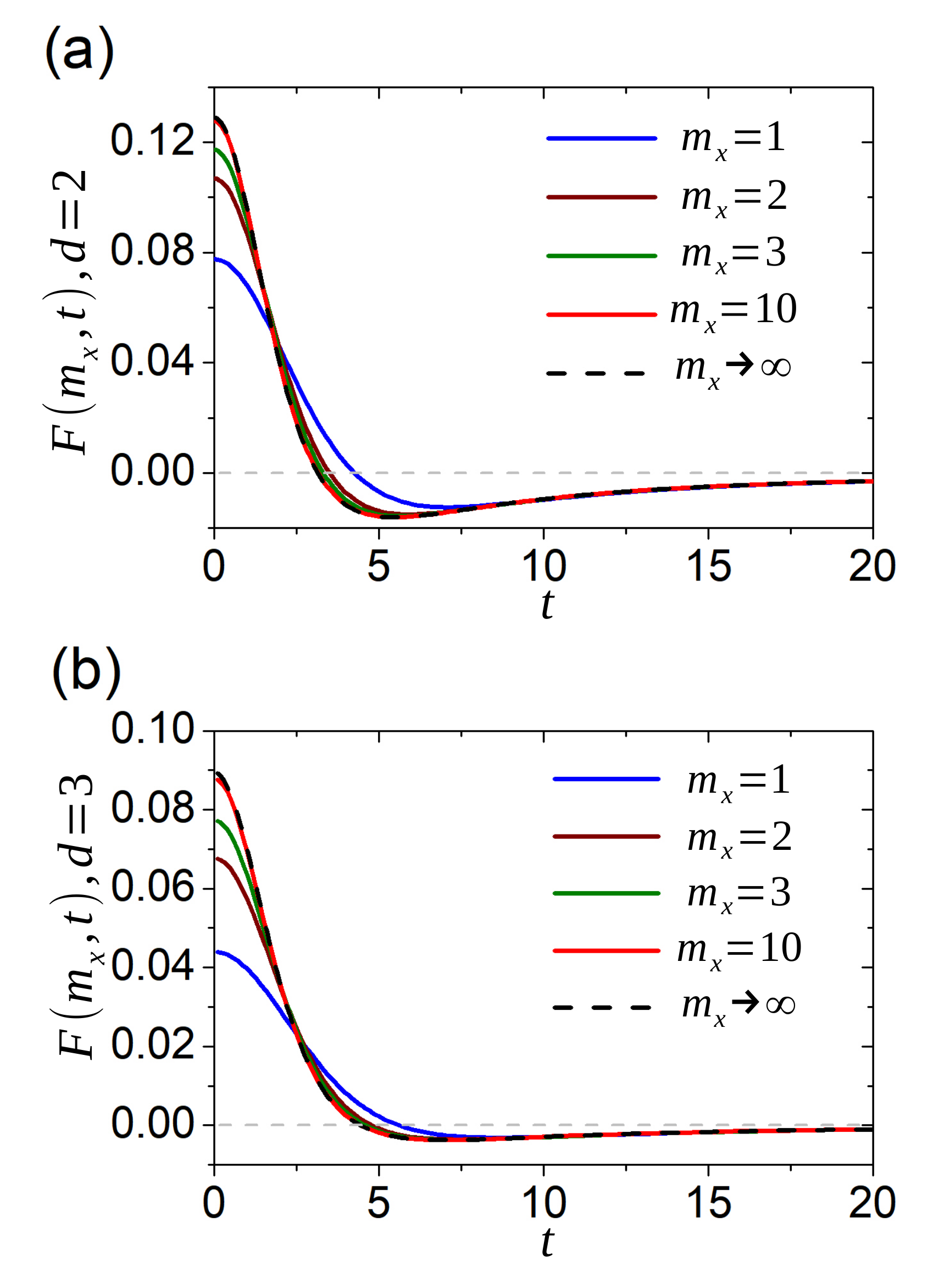}
\caption{Fourier coefficients Eqs.~(\ref{Fgmq}) and (\ref{FWmq1}) for the second order correction to conductance in the limit $W/L\gg 1$ in two dimensions (a) and three dimensions (b), respectively. Solid curves correspond to $m_x=1$, $2$, $3$, and $10$; dashed curves describe the thermodynamic
limit $m_x \rightarrow\infty$ calculated from Eqs.~(\ref{sigma2g}) and (\ref{sigma2W}).
\label{fig:pert}}
\end{figure}

In Fig.~\ref{fig:pert} we show the Fourier coefficient $F$ as a function of $t = 
\pi \bar q_{\perp}/q_x$ and for various values of $m_x = q_x L/\pi$. The figure,
as well as the asymptotic expressions Eqs.~(\ref{sigma2g}) and (\ref{sigma2W}) for the
limit $m_x \to \infty$, shows that $F$ is positive for small $t$ and becomes negative if $t$ is sufficiently large. It 
follows that, {\em a priori}, there is no definite sign for the correction $g^{(2)}$ to
the dimensionless conductance. In particular, a well-chosen periodic ``disorder 
potential'' $U(\vr) \propto\sin (q_{x}x) \cos(q_{y}y)$ in two dimensions results in a 
negative correction $g^{(2)}$ if $v_{y}q_{y}/v_{x}q_{x}$ is sufficiently
large. A similar conclusion applies to $d=3$. 
Finding the magnitude and sign of the conductance correction for a generic 
disorder potential requires a statistical analysis involving the sum of a large number 
of Fourier components, which is performed in the next Section.

\section{Conductance fluctuations\label{sec:minimal}}

We now apply the general results of Sec.~\ref{sec:pert} to a random potential
$U(\vr)$. We take $U(\vr)$ to have a Gaussian distribution with zero mean
and with two-point correlation function
\begin{align}
  \langle U(\vr)  U(\vr') \rangle =&\,
  \frac{\hbar^2 K}{\xi^2 (2 \pi)^{d/2}}
  e^{-|\vr-\vr'|^{2}/2\xi^{2}} \nonumber \\ &\, \mbox{} \times
    \left\{ \begin{array}{ll}
    v_x v_y, & d=2, \\
    v_y v_z, & d=3,
  \end{array} \right. \label{Urandomsr}%
\end{align}
where $K$ is the dimensionless disorder strength and $\xi$ is the correlation 
length\footnote{One can absorb an arbitrary dimensionless function of 
$v_y/v_x$ and $v_z/v_x$ into $K$, but the scaling of cumulants $\kappa_n$ with $n\geq 2$ is simplified by the convention in 
Eq.~(\ref{Urandomsr}).}. Disorder potentials of this form have frequently been used in 
numerical simulations of the conductance of disordered Dirac and Weyl nodes, see,
{\em e.g.}, Refs.~\onlinecite{PhysRevLett.99.106801,PhysRevLett.113.026602}.

Upon Fourier transforming the short-range Gaussian disorder model Eq.~(\ref{Urandomsr}) according to Eq.~(\ref{Uxyg}) we find
\begin{equation}
  U(\vq) = r(\vq) e^{-q^2 \xi^2/4}
  \sqrt{\frac{K \hbar^2}{L W^{d-1}}}
  \left\{ \begin{array}{ll}
    \sqrt{v_x v_y} , & d=2, \\
    \sqrt{v_y v_z}, & d=3,
  \end{array} \right. \label{randomU}%
\end{equation}
where the $r(q_x,\vq_{\perp}) = r(q_x,-\vq_{\perp})^*$ are (otherwise) mutually independent standard normal variates each obeying a zero-mean normal distribution with variance 
\begin{equation}
  \langle |r(q_x,\vq_{\perp})|^2 \rangle= 2 \left(1 - \frac{1}{2} \delta_{\vq_{\perp},0} \right).
\end{equation}

\begin{figure}
\centering
\includegraphics[width=0.5\textwidth]{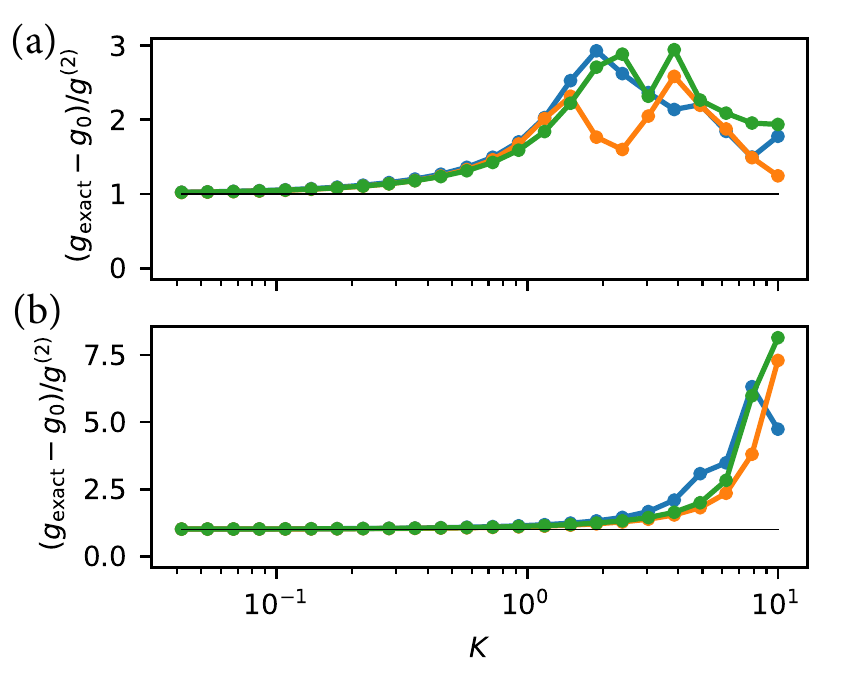}
\caption{Comparison of the numerically evaluated exact disorder-induced correction $g_{\text{exact}}-g_0$ to the dimensionless conductance and the same quantity $g^{(2)}$ in second-order perturbation theory. Each curve corresponds to a single realization of the random potential $U(\vr)$ according to Eq.~(\ref{randomU}) with $K=1$, rescaled by a factor $\sqrt{K}$ to reveal the dependence on the disorder strength. Panel (a) is for an isotropic two-dimensional Dirac node with $L/\xi=10$ and $W/L=10$; panel (b) is for an isotropic three-dimensional Weyl node with $L/\xi=6$ and $W/L=5$. Numerical calculations were performed using the approach of Ref.~\onlinecite{PhysRevLett.99.106801}; perturbation theory results are given by Eqs.~(\ref{G2g}) and (\ref{Fgmq}) for $d=2$ (a) and Eqs.~(\ref{G2W}) and (\ref{FWmq1}) for $d=3$ (b).\label{fig:pertcompare}}
\end{figure}

We first check the range of validity of our perturbation theory results by randomly generating a disorder potential profile from Eq.~(\ref{randomU}) with $K=1$ and subsequently rescale the random potential with $\sqrt{K}$. A comparison between the second-order result of Eqs.~(\ref{G2g}), (\ref{Fgmq}), (\ref{G2W}), and (\ref{FWmq1}) and a numerically exact calculation using the method of Ref.~\onlinecite{PhysRevLett.99.106801} is shown in Fig.~\ref{fig:pertcompare}. For all disorder realizations we have generated, the quadratic scaling of $g^{(2)}$ with the strength of the disorder potential holds for $K \lesssim 0.1$ in a two-dimensional system with $L/\xi=10$, and $K \lesssim 1$ in a three-dimensional system with $L/\xi=6$. In two dimensions, we find that the range of validity of our second-order perturbation theory shrinks as $L/\xi$ grows (data not shown); in other words, for a given $K$, higher order corrections become progressively more important for larger $L/\xi$. This is consistent with the renormalization group analysis of the Gaussian disorder in graphene \cite{PhysRevB.74.235443,PhysRevB.79.075405}: the leading fourth-order or $O(K^2)$ correction depends logarithmically on the infrared cutoff (in this case the length $L$), and sufficiently large $L/\xi$ eventually drives the system away from the quasi-ballistic regime into the diffusive regime. On the other hand, in three dimensions, the disorder potential is an irrelevant perturbation \cite{PhysRevB.33.3257,PhysRevLett.107.196803,PhysRevB.84.235126,PhysRevLett.114.166601,PhysRevB.94.220201} and its strength scales as $\xi/L$; thus it is possible that our perturbation theory applies for $K \lesssim 1$ even in the thermodynamic limit.

We now turn to the cumulants of the probability distribution of $g^{(2)}$. Inserting Eq.~(\ref{randomU}) into Eq.~(\ref{G2g}) or (\ref{G2W}), we have%
\begin{equation}
  g^{\left(2\right)}= \frac{1}{2} \sum_{\vq} 
  V(\vq) |r(\vq)|^2,\label{randomg2}
\end{equation}
where 
\begin{align}
  V(\vq) &=\frac{K\xi^{d-2}}{2L^{d}}
  \frac{\pi^{2}}{q_{x}^{2}}  e^{-q^2 \xi^2/2}
  F\left(  \frac{q_{x}L}{\pi},\frac{\pi v_{y}q_{y}}{v_{x}q_{x}},\frac{\pi v_{z}q_{z}}{v_{x}q_{x}}\right) \text{,}
\end{align}
understanding that the (third) $q_{z}$ argument is absent in two dimensions. It is now
straightforward to find the first few cumulants of the probability distribution of $g^{\left(  2\right)
}$,
\begin{subequations}
\label{cumulants}
\begin{align}
  \kappa_{1} \equiv&\, \langle g^{\left(  2\right)  }\rangle 
  \nonumber \\ =&\, 
  \sum_{\vq} V(\vq) \left(1 - \frac{1}{2} \delta_{\vq_{\perp},0} \right),
  \label{eq:kappa1} \\
  \kappa_{2}\equiv &\, \langle (  g^{(2)}-\langle g^{(2)} \rangle )^2 \rangle
  \nonumber \\ =&\,
  2 \sum_{\vq} V(\vq)^2 \left(1 - \frac{1}{2} \delta_{\vq_{\perp},0} \right),
  \label{eq:kappa2} \\
  \kappa_{3} \equiv&\, \langle (  g^{(2)}-\langle g^{(2)}\rangle)^{3}\rangle 
  \nonumber \\ =&\,
  8 \sum_{\vq} V(\vq)^{3} \left(1 - \frac{1}{2} \delta_{\vq_{\perp},0} \right),
  \label{eq:kappa3} \\
  \kappa_{4} \equiv&\, \langle (  g^{(2)}-\langle g^{(2)}\rangle)^{4}\rangle 
  - 3 \langle (g^{(2)}-\langle g^{(2)}\rangle)^{2}\rangle^{2}\nonumber \\
  =&\, 48 \sum_{\vq} V(\vq)^{4} \left(1 - \frac{1}{2} \delta_{\vq_{\perp},0} \right). \label{eq:kappa4}
\end{align}
\end{subequations}
Below we evaluate these expressions in the limit $\xi\ll L$ that the correlation length of the
disorder potential is much smaller than the sample length. We also take the limit of large
aspect ratio, $W/L\gg1$, which allows us to replace the summations over $q_{y}$ and $q_{z}$ by
integrals, while keeping $q_{x}$ finite.

\subsection{Dirac node}

In the case of the two-dimensional Dirac node, the expectation value $\langle g^{(2)}\rangle $ is 
most easily calculated starting from the
real-space expression Eq.~(\ref{OUsqcont}). Due to the short-range
correlations, only $\left\vert \mathbf{r}-\mathbf{r}^{\prime}\right\vert
\lesssim\xi$ contributes significantly, so we can Taylor-expand some of the
cosines and hyperbolic cosines in the integrand and neglect the $x+x^{\prime}$
term entirely:%
\begin{align}
\langle g^{\left(  2\right)  }\rangle  \approx &\, 
  \frac{Wv_{x}^{2}}{\pi^2 \hbar^{2}v_{y}^{4}L^{2}}\int_{0}^{L}dxdx^{\prime}
  \int d\delta y\frac{K\hbar^{2}v_{x}v_{y}}{2\pi\xi^{2}}\nonumber\\
  & \mbox{} \times 
  \frac{v_y^2 \delta y^{2}}{v_{x}^{2} \delta y^{2}+ v_y^2 (x-x')^{2}}
  e^{-[(x-x')^{2}+\delta y^{2}]/2\xi^{2}}
  \nonumber\\
  \approx&\, \frac{W}{L}\frac{K}{\pi^{2}}
  \frac{v_{x}^2}{v_{x} v_{y}+v_{y}^{2} } \text{.}\label{g2avg}%
\end{align}
In the second line we have switched to polar coordinates and performed the radial and angular 
integrals separately.

The same result can also be obtained using the Fourier representation Eq.~(\ref{G2g}) of $g^{(2)}$. We briefly discuss this calculation, too, as it requires some care and because we will need the Fourier representation to calculate the higher cumulants of $g^{(2)}$ and to calculate $g^{(2)}$ in the case of the Weyl node. We first notice that it is possible to replace the coefficient $F$ in Eq.~(\ref{G2g}) with its asymptotic expression Eq.~(\ref{sigma2g}), so that in the thermodynamic limit $W,L\to \infty$,
\begin{align*}
  \langle g^{\left(  2\right)  }\rangle \approx & \frac{2KW}{\pi L^2}
  \sum_{q_x} 
  \int_{-\infty}^{\infty}\frac{dq_{y}}{2\pi} 
  \frac{v_x^2 (v_x^2 q_x^2 - v_y^2 q_y^2)}{(v_x^2 q_x^2 + v_y^2 q_y^2)^2}
  e^{-q^{2}\xi^{2}/2}.
\end{align*}
The difference between $F$ and its asymptotic approximation Eq.~(\ref{sigma2g}) is significant only when $q_x L/\pi \sim 1$ and $\pi v_y q_y/v_x q_x \sim 1$, but even in this case Eq.~(\ref{sigma2g}) correctly estimates the order of magnitude of $F$, as shown in Fig.~\ref{fig:pert}. Because the dominant contribution to $\langle g^{(2)}\rangle $ comes from $q_x \lesssim 1/\xi$ and $q_y \lesssim 1/\xi$, we expect the error caused by the replacement to be $O(\xi^2/L^2)$.

It might be tempting at this point to replace the summation over $q_x$ by an integral and perform the integrals over $q_x$ and $q_y$ together in polar coordinates. However, this ignores the fundamental anisotropy of the large-aspect ratio limit $W/L\gg 1$, which requires taking the limit $W\to \infty$ first and then the limit $L\to \infty$; thus $q_y$ is effectively continuous, whereas $q_x$ is not, and one need to calculate the $q_y$ integral under the assumption of a nonzero $q_x$. This is achieved by the substitution $\zeta = v_{y}q_{y}/v_{x}q_{x}$. Integrating by parts, we find
\begin{align}
\langle g^{\left(  2\right)  }\rangle  \approx&\, \frac{KW}{\pi^{2} L^2}
  \frac{v_{x}^{3}}{v_{y}^{3}} \sum_{q_x} q_{x}\xi^{2}e^{-q_{x}^{2}\xi^{2}/2}\nonumber\\
  & \times\int_{-\infty}^{\infty}d\zeta
  \frac{\zeta^{2}}{1+\zeta^{2}}
  e^{-v_{x}^{2} \zeta^{2}q_{x}^{2}\xi^{2}/2 v_y^2}.
\end{align}
At this point we may replace the summation over $q_x$ by an integral and perform the $q_x$-integral, before the $\zeta$-integral. This reproduces Eq.~(\ref{g2avg}).

\begin{figure}
[h]\centering\includegraphics[width=0.5\textwidth]{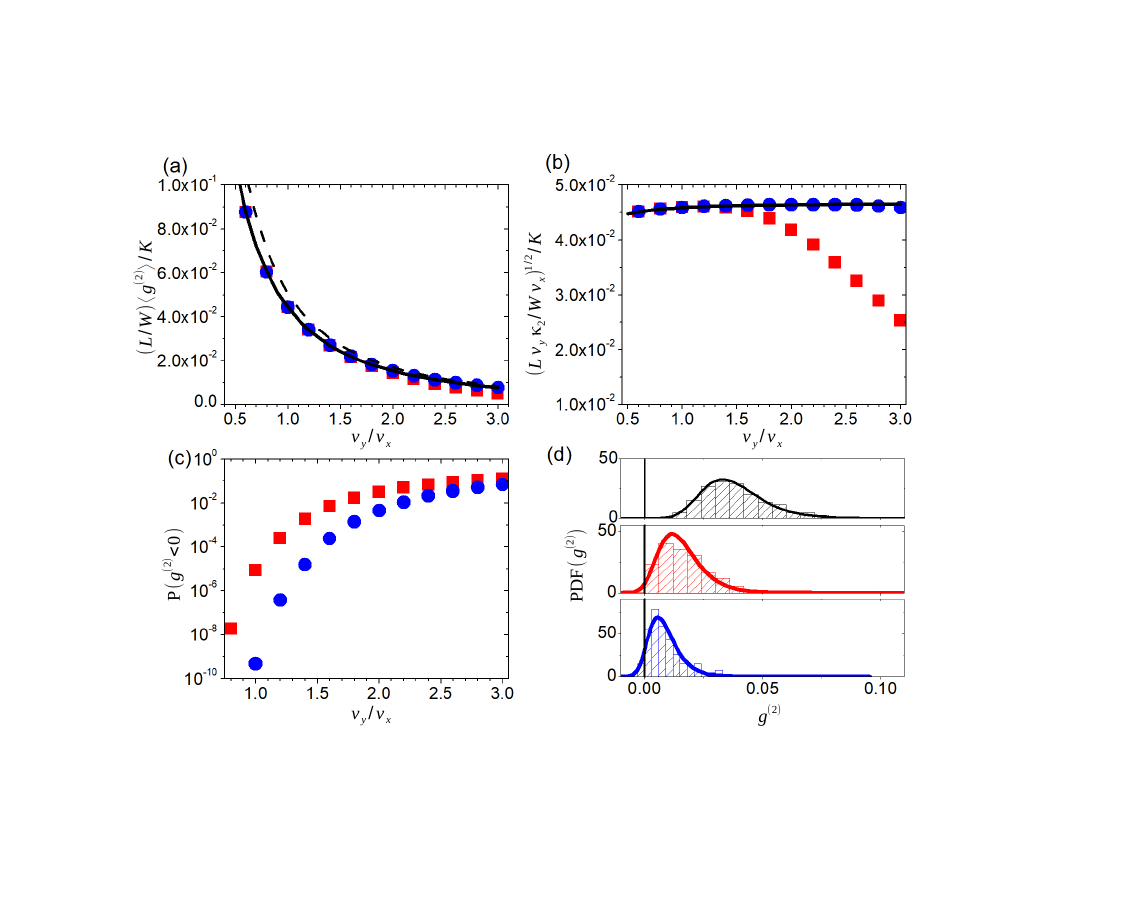}
\caption{
(a) and (b): Normalized expectation value $( L/W)\langle g^{( 2)} \rangle /K$ (a) and normalized standard deviation $(Lv_y \kappa_{2}/Wv_x)^{1/2} /K$ (b) of the second-order disorder-induced conductance correction $g^{( 2)}$ as functions of $v_y/v_x$ for an anisotropic Dirac node with an isotropic Gaussian random potential of dimensionless strength $K$ and correlation length $\xi$. The data points correspond to $W/L=10$ (red squares) or $20$ (blue circles) and $L/\xi=20$; the solid lines represent the limit $W/L\to\infty$ at $L/\xi = 20$. The dashed black line in panel (a) represents the simultaneous limit $W/L$, $L/\xi \to \infty$ of Eq.~(\ref{g2avg}).
(c): Probability that $g^{(2)}<0$ as a function of $v_y/v_x$ for $L/\xi=20$, $W/L=10$ (red squares) and $W/L=20$ (blue circles).
(d): Probability distribution of $g^{\left( 2 \right)}$ for $K=0.1$, $L/\xi=10$, $W/L=10$ and $v_y/v_x=1$, $1.8$ and $2.4$ (top to bottom), calculated by Fourier transforming the characteristic function from the perturbation theory. 
Also shown are the normalized histograms of the disorder correction to conductance obtained from the scattering matrix approach with $919$, $3593$ and $5480$ disorder realizations respectively. For reference, $g^{(2)}=0$ is marked by vertical lines.
\label{fig:distg}}
\end{figure}

For the isotropic dispersion $v_x=v_y$, one recovers the result of Refs.~\onlinecite{PhysRevB.79.075405,PhysRevB.82.085419} that $\langle g^{\left(  2\right)  }\rangle =(W/L)K/(2\pi^2)$. Here the weak anti-localization correction to bulk conductivity seen in the numerical study of Ref.~\onlinecite{PhysRevLett.99.106801} and responsible for the scaling flow to the diffusive regime \cite{PhysRevB.82.085419} in the limit $L/\xi \to \infty$ is absent, since we have limited ourselves to the ballistic regime by considering no more than two scattering events. Figure~\ref{fig:distg}a shows $\left( L/W \right)\langle g^{\left( 2 \right)} \rangle /K$ as a function of the dispersion anisotropy $v_y/v_x$. The average disorder correction $\langle g^{(2)} \rangle$ approaches the limiting value for $W/L\gg 1$ and $L/\xi \gg 1$ from below as $W/L$ or $L/\xi$ increases, respectively. For a fixed aspect ratio $W/L$, the average correction deviates from the limiting value at $W/L\gg 1$ more strongly for larger values of $v_y/v_x$. 

The cumulants $\kappa_n$ with $n \ge 2$ behave quite differently from the mean: The dominant contributions to these cumulants comes from Fourier components with small $q_{x}$, {\em i.e.} with longitudinal wavelengths comparable to the length $L$. To see this, we note that, after replacing the summation over $q_y$ by an integral and performing the integration, the summand in Eqs.\ (\ref{eq:kappa2})--(\ref{eq:kappa4}) diverges for small $q_x$ and is proportional to $W L^{-2n} q_x^{1-2n} v_x/v_y$, so that $\kappa_n \propto W v_x/L v_y$ if $n \ge 2$. To verify this scaling, in Fig.~\ref{fig:distg}b we show the quantity $(L v_y \kappa_2/W v_x)^{1/2}$ as a function of the ratio $v_y/v_x$ for two different values of the aspect ratio $W/L$ and also in the limit $W/L\gg 1$. As expected $(L v_y \kappa_2/W v_x)^{1/2} $ is almost independent of $v_y/v_x$ in the limit $W/L \gg 1$. In the limit $L/\xi \to \infty$, we find $(L v_y \kappa_2/W v_x)^{1/2}/K \approx 0.0484$. The increase with $K$ for $K \lesssim 1$ is consistent with numerical simulations of Ref.\ \onlinecite{EurophysLett.79.57003}.

Deviations from a Gaussian form of the probability distribution of $g^{(2)}$ are characterized by the skewness $\kappa_3/\kappa_2^{3/2}$ and the excess kurtosis $\kappa_4/\kappa_2^2$, which scale proportional to $(L v_y/W v_x)^{1/2}$ and $(L v_y/W v_x)$, respectively. The fact that the skewness and the excess kurtosis are suppressed by powers of $L/W$ suggests the distribution of $g^{(2)}$ approaches a normal distribution in the limit $W/L\gg 1$. This provides an estimate of the probability to find a negative disorder-induced conductance correction,
\begin{align}
P(g^{(2)}<0)  \approx&\, \frac{1}{2}\operatorname{erfc}\left(  \frac{\kappa_{1}}{\sqrt{2\kappa_{2}}}\right)\nonumber\\
 \approx &\, \frac{1}{2} \operatorname{erfc}\left[ c\left( \frac{L}{\xi} \right)\sqrt{\frac{W}{L}\frac{v_x}{v_y}}\frac{v_x}{v_x+v_y}\right],
  \label{eq:Pnegative}
\end{align}
where $\operatorname{erfc}(x)$ is the complementary error function with an asymptotic expansion $\operatorname{erfc}(x)\sim e^{-x^2}/\sqrt{\pi}x$ in the limit $x\to \infty$ and $c(L/\xi)$ is a function that depends weakly on its argument, taking the value $c \approx 1.48$ for $L/\xi \to \infty$. We conclude that the probability of disorder reducing the conductance is exponentially suppressed as a function of the aspect ratio $W/L$ in the limit $W/L\rightarrow\infty$, although it increases rapidly as a function of $v_y/v_x$. Figure \ref{fig:distg}c shows $P\left( g^{\left(  2\right)  }<0\right)  $ as a function of $v_y/v_x$ for $L/\xi=20$ and different aspect ratios $W/L$. While the probability for $g^{(2)}<0$ is vanishingly small for an isotropic dispersion --- $P(g^{(2)} < 0) \sim 10^{-5}$ when $W/L=10$ --- we also see that it can become significantly larger when $v_{y}/v_{x}$ increases, becoming of order of $0.1$ when $v_y/v_x \gtrsim 3$. In other words, it becomes more likely for isotropic disorder to reduce the conductance below the quasi-ballistic value (\ref{eq:Gmin2anis}) as the Dirac cone is compressed in the transverse direction.

This is further visualized in Fig.~\ref{fig:distg}d, where we plot the probability distribution of $g$ for $L/\xi=10$, $W/L=10$, $K=0.1$ and various values of the anisotropy parameter $v_y/v_x$, accompanied by the normalized histograms of numerical data from the scattering matrix approach. The probability density function is calculated by (numerically) Fourier transforming the characteristic function of $g^{(2)}$, which can be found exactly because Eq.~(\ref{randomg2}) is quadratic in the normal variates $r$. 

\subsection{Weyl node}

We now turn to the case of three dimensions. For simplicity we assume that $v_y=v_z\equiv v_{\perp}$. As in the two-dimensional case, in the limit $L/\xi$, $W/L \gg 1$ we can approximate the Fourier coefficient $F$ by Eq.~(\ref{sigma2W}), yielding
\begin{align}
\langle g^{\left(  2\right)  }\rangle  \approx& \frac{W^{2} K \xi \ln 2}{\pi L^{3}}
  \sum_{q_x} 
  \int_{-\infty}^{\infty}\frac{d \vq_{\perp}}{(2\pi)^{2}}\nonumber\\
& \times 
  \frac{v_x^2(2 v_x^2 q_x^2 - v_{\perp}^2 q_{\perp}^2)}{(v_x^2 q_x^2 + v_{\perp}^2 q_{\perp}^2)^2}
  e^{-q^{2}\xi^{2}/2},
\end{align}
where $\vq_{\perp} = q_y \ve_y + q_z \ve_z$. Proceeding analogously to the two-dimensional case, we make the substitution $\zeta = v_{\perp} q_\perp / v_x q_x$ and switch to polar coordinates for the transverse momentum $\vq_{\perp}$. The sum over $q_x$ can then be written as a Jacobi $\vartheta$ function, which is subsequently expanded in an asymptotic series in $\xi/L$. \cite{JMathPhys.58.011702} This leads to
\begin{align}
\langle g^{\left(  2\right)  }\rangle  \approx&\, \frac{W^{2} K \ln 2}{\pi L^{2}} \frac{v_{x}^{2}}{v_{\perp}^{2}}
\int_{0}^{\infty}\frac{d\zeta}{4\pi}\frac{2-\zeta}{\left(  1+\zeta\right)  ^{2}}\nonumber\\
& \times \left[ \sqrt{\frac{v_{\perp}^2}{2 \pi(v_{\perp}^2 + \zeta v_x^2)}} -\frac{\xi}{2L}+O\left( \frac{\xi^2}{L^2}\right)\right].
\label{g2avgW0}
\end{align}
The $\zeta$ integral is convergent for the first term in the square brackets. For the second term, the integral should be cut off at $\zeta\sim v_\perp^2 L^2/v_x^2 \xi^2$, which is where the asymptotic expansion fails. The result is
\begin{align}
  \label{g2avgW}
  \langle g^{(2)}\rangle \approx &\,
  \frac{W^2 K \ln 2}{8 \pi^2 L^2}
  \left[
   \frac{2 \theta[2+\cos (2 \theta)] - 3 \sin(2 \theta)}{(2 \pi)^{1/2} \sin^3 \theta \cos \theta} 
  \right. \nonumber \\ &\, \left. \mbox{} 
  + \frac{\xi}{L} \frac{1}{\cos^2 \theta} \left( c' + 2 \ln \frac{L\cos \theta}{\xi} \right) \right],
\end{align}
where $\cos \theta = v_{\perp}/v_x$ and $c'$ is a number of order unity that cannot be determined from the approximation in Eq.~(\ref{g2avgW0}), but which weakly depends on $L/\xi$ ($c' \approx 1.0$ for $L/\xi = 6 \times 10^2$, $c' \approx 0.7$ for $L/\xi = 2.4 \times 10^4$). The second line in Eq.\ (\ref{g2avgW}) is a small correction to $\langle g^{(2)} \rangle$ that goes to zero in the limit $L/\xi \to \infty$.

\begin{figure}
\centering
\includegraphics[width=0.5\textwidth]{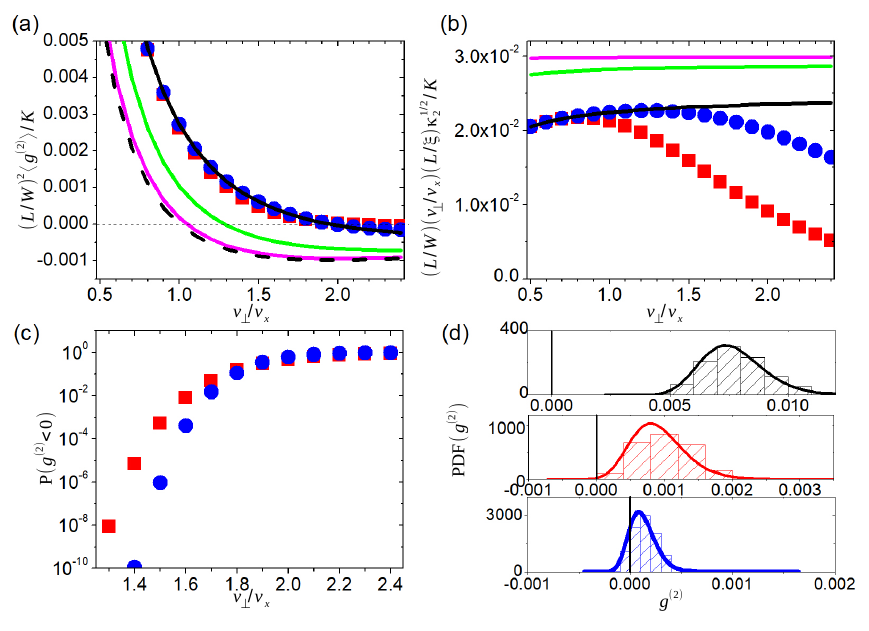}
\caption{
(a) and (b): Expectation value $\langle g^{(2 )} \rangle$ (a) and standard deviation $\kappa_2^{1/2}$ (b) of the disorder-induced conductance correction as functions of $v_{\perp}/v_x$ for an anisotropic Weyl node with isotropic short-range disorder. The data points correspond to $W/L=5$ (red squares) or $8$ (blue circles) and $L/\xi=12$; the solid lines correspond to $W/L\gg 1$, $L/\xi=12$ (black), $60$ (green) and $600$ (magenta). The dashed black line in panel (a) represents the limit $W/L$, $L/\xi \gg 1$ of Eq.~(\ref{g2avgW}).
(c): Probability that $g^{( 2)}<0$ as a function of $v_{\perp}/v_x$ for $L/\xi=12$, $W/L=5$ (red squares) and $W/L=8$ (blue circles).
(d): Probability density functions of $g^{( 2)}$ for $K=0.1$, $L/\xi=6$, $W/L=5$ and $v_y/v_x=1$, $1.8$ and $2.4$ (top to bottom), overlaid with the normalized histograms of the disorder correction to conductance obtained from the scattering matrix approach with $3100$, $521$ and $458$ disorder realizations respectively. The reference point $g^{\left( 2 \right)}=0$ is marked by vertical lines.
\label{fig:distW}}
\end{figure}

The result Eq.~(\ref{g2avgW}) for the limit $W/L$, $L/\xi \gg 1$ is shown in Fig.~\ref{fig:distW}a, together with a numerical evaluation of the disorder average of the exact result Eq.~(\ref{G2W}) for $\langle g^{(2)} \rangle$ for finite $W/L$ and $L/\xi$, as well as for $W/L \to \infty$ while keeping $L/\xi$ finite. For an isotropic dispersion, $v_{\perp} = v_{x}$, corresponding to $\theta = 0$ in Eq.\ (\ref{g2avgW}), the disorder average $\langle g^{(2)} \rangle$ vanishes in the $L/\xi \to \infty$ limit. This is consistent with the fact that short-range disorder is an irrelevant perturbation for a Weyl semimetal with chemical potential at the nodal point. \cite{PhysRevB.84.235126,PhysRevLett.114.166601} On the other hand, for an anisotropic Weyl cone, isotropic disorder renormalizes the effective anisotropy $v_\perp/v_x$, which enters into the expression for the quasi-ballistic conductance Eq.~(\ref{eq:Gmin3anis}), which explains why $\langle g^{(2) }\rangle$ is negative for $v_\perp/v_x > 1$ and positive for $v_{\perp}/v_x < 1$ in the limit $L/\xi \to \infty$, see Fig.\ \ref{fig:distW}a. For finite $L/\xi$, $\langle g^{(2) }\rangle$ only becomes negative when $v_\perp/v_x$ exceeds a threshold value larger than unity.

Analogous to the two-dimensional case, we find that the cumulants $\kappa_n$ with $n \ge 2$ are controlled by the aspect ratio and the anisotropy: the summand in Eqs.~(\ref{eq:kappa2})--(\ref{eq:kappa4}) diverges for small $q_x$ and is proportional to $W^2 \xi^n L^{-3n} q_x^{2-2n} v_x^2/v_\perp^2$, so that $\kappa_n \propto W^2 \xi^n v_x^2/L^{2n+2} v_\perp^2$ if $n \ge 2$. In contrast to the two-dimensional case, the cumulants $\kappa_n$ have an additional smallness $\propto (\xi/L)^n$ if $L/\xi \gg 1$. For the variance $\kappa_{2}$, this scaling behavior is confirmed in Fig.~\ref{fig:distW}b. In the limit $W/L\gg 1$ and $L/\xi \to \infty$, we find that $\left( L^2 v_{\perp}/ \xi W v_x \right) \kappa_{2}^{1/2}/K \approx 0.030$, independent of $v_{\perp}/v_x$. The suppression of conductance fluctuations for $L/\xi \gg 1$ is consistent with the expectation that disorder is an irrelevant perturbation at the Weyl point.

In the limit of $W/L\gg 1$, the skewness $\kappa_3/\kappa_2^{3/2}$ and the excess kurtosis $\kappa_4/\kappa_2^2$ are proportional to $(  L/W)  (  v_{\perp}/v_{x})$ and $(  L/W)  ^{2} (  v_{\perp}/v_{x})^{2}$, respectively, so the distribution of $g^{(2)}$ again approaches a normal distribution. In particular, when $v_\perp=v_x$, using Eq.~(\ref{g2avgW}) we can estimate the probability to find a negative disorder-induced conductance correction as
\begin{align}
P(g^{(2)}<0)  \sim&\, \frac{1}{2}\operatorname*{erfc}\left[  0.207\frac{W}{L}\left(  c^{\prime} + 2\ln\frac{L}{\xi}\right)  \right].
  \label{eq:PnegativeW}
\end{align}
Thus for an isotropic Weyl cone, the probability of disorder reducing the conductance is exponentially suppressed as a function of $W^2/L^2$ in the limit $W/L\rightarrow\infty$.

Since $\langle g^{(2) }\rangle$ is nonzero if the dispersion is not isotropic, the probability to find a negative disorder-induced conductance correction strongly depends on $v_{\perp}/v_x$: It is significantly smaller than the estimate (\ref{eq:PnegativeW}) if $v_{\perp}/v_x < 1$, whereas $P(g^{(2)} < 0) \to 1$ in the limit $W/L$, $L/\xi \to \infty$ if $v_{\perp}/v_x > 1$. Figure \ref{fig:distW}c shows $P(  g^{(  2)  }<0)$ versus $v_{\perp}/v_x$ for $L/\xi=12$ and different $W/L$, clearly confirming the strong dependence on $v_{\perp}/v_x$ for large aspect ratios $W/L$. The increase of $P(g^{(2)}<0)$ with $v_{\perp}/v_x$ is also illustrated in Fig.~\ref{fig:distW}d, where we plot the probability density functions of $g^{(  2)  }$ for $L/\xi=6$, $W/L=5$ and various values of $v_{\perp}/v_x$. As in the case of a two-dimensional Dirac node, isotropic disorder has a larger probability to reduce the conductance relative to the quasi-ballistic value as the Weyl cone is compressed in the transverse directions.

\section{Discussions and conclusions\label{sec:conclusion}}

In this work, we have addressed the mesoscopic transport of an anisotropic Dirac node in a two-dimensional electron gas or a Weyl node in a three-dimensional semimetal. We calculated the conductance for Fermi energy at the nodal point to second order in a perturbing potential and evaluated the statistics of the conductance for a generic model of short-range disorder. Our theory is controlled close to the quasi-ballistic regime in which wavepackets scatter only a few times in the sample. 

In two dimensions, the conductance is normally distributed if the aspect ratio $W/L$ is large, with a variance that scales as $W v_{x}/L v_{y}$, where $v_{y}/v_x$ is a measure of the anisotropy of the dispersion at the nodal point. Isotropic short-range disorder always increases the conductance on the average. Because conductance fluctuations are small for large $W/L$, disorder fluctuations for which the disorder-induced conductance correction is negative are extremely rare (although they do occur). This explains the empirical observation of the absence of such disorder realizations even in large-scale numerical simulations.

In three dimensions, disorder is an irrelevant perturbation in the renormalization-group sense. \cite{PhysRevB.33.3257,PhysRevLett.107.196803,PhysRevB.84.235126,PhysRevLett.114.166601,PhysRevB.94.220201} Our perturbative calculation of the conductance distribution is consistent with this observation, but also further refines it. In particular, we find that isotropic short-range disorder affects the mean conductance if the dispersion is anisotropic --- a finding that may be interpreted as a disorder-induced renormalization of the dispersion anisotropy. In particular, if $v_{\perp}/v_x > 1$ (Fermi velocity in the direction of current flow is smaller than the Fermi velocity transverse to the direction of current flow) disorder decreases the conductance on the average, whereas the average disorder-induced correction is positive if $v_{\perp}/v_x < 1$. 
The conductance fluctuations, however, are proportional to $(v_x/v_{\perp})^2 (W/L)^2 (\xi/L)^2$, which at large $L/\xi$ has an additional suppression compared to the naive generalization of the two-dimensional result. It is the absence of disorder-induced conductance fluctuations in the limit $L/\xi \to 1$ that makes these findings, again, consistent with the expectation from scaling theory that disorder is an irrelevant perturbation in this case.

Finally, we would like to make a few comments on the experimental relevance of this work. Although our results were obtained in the context of graphene (in two dimensions) or a Weyl semimetal (in three dimensions) with short-range disorder, our perturbative calculation of the conductance applies to a Dirac spectrum with an arbitrary scattering potential. Hence, our results may also prove useful in describing nodal-point transport in engineered mesoscopic structures such as superlattices \cite{NatPhys.8.382} and electrostatic confinement potentials, \cite{NatPhys.12.1032} as long as the potential strength is sufficiently small that the quasi-ballistic assumption holds.

Although quite a large number of measurements of mesoscopic conductance fluctuations of graphene \cite{Science.312.1191,PhysRevLett.97.016801,Nature.446.56,PhysRevB.77.155429,PhysRevLett.102.066801,PhysRevLett.104.186802,PhysRevB.86.161405,PhysRevLett.109.196601,SciRep.6.33118,PhysRevB.95.125427} and Dirac/Weyl semimetal devices \cite{PhysRevB.94.161402} have been reported in the literature, a direct comparison of these results with our theoretical predictions is not possible. The reason is that a measurement of the conductance fluctuations needs a means to generate an ensemble of (effectively) different disorder realizations. Experimentally, this is achieved by considering variations of gate voltage or a magnetic field, relying on the ergodic hypothesis, as in the case of conventional two- or three-dimensional conductors. \cite{PhysRevB.35.1039} However, this is not a viable approach to address the distribution of the mimimal conductance, which requires tuning of the gate voltage to the nodal point and zero magnetic field. (Indeed, the ergodic hypothesis is seen to break down in graphene around the Dirac point, \cite{PhysRevLett.104.186802,PhysRevB.86.161405,SciRep.6.33118} whereas the application of a magnetic field in Dirac/Weyl semimetals can open up an excitation gap at the nodal point. \cite{PhysRevB.85.195320}) The motion of impurities associated with thermal cycling \cite{JPhysIFrance.2.357} or low-frequency noise \cite{PhysRevLett.56.1960,PhysRevLett.58.1240} would provide an alternative method to obtain a disorder ensemble that can be used to measure the conductance distribution at the nodal point.

\begin{acknowledgments}
We thank Sergey Syzranov for useful discussions. We acknowledge support by project A02 of the CRC-TR 183 (PWB, ZS) and by the German National Academy of Sciences Leopoldina through grant LPDS 2018-12 (BS).
\end{acknowledgments}

\bibliography{minimalG}

\end{document}